\documentstyle[12pt]{article} \begin{document}

%--------------------------------------

\newcommand{\be}{\begin{equation}} \newcommand{\ee}{\end{equation}}
\newcommand{\bea}{\begin{eqnarray}} \newcommand{\eea}{\end{eqnarray}}

\newcommand{\Uno}{{\rm 1\hspace{-0.9mm}l\hspace{-1.0mm}}} \newcommand{\nat}{{\rm
I\hspace{-0.4mm}l\hspace{-1.1mm}N}}
\newcommand{\entero}{\rule[2.2mm]{0.1mm}{0.5mm}\hspace{-0.3mm}{\sf Z}\hspace{-1.7mm}{\sf
Z}\hspace{-0.3mm}\rule{0.1mm}{0.5mm}} \newcommand{\rea}{{\rm I\hspace{-0.8mm}R}}
\newcommand{\comp}{\mbox{\rm\hspace{1.2mm}\rule[0.05mm]{0.1mm}{2.5mm}\hspace{-1.2mm}C}}
\newcommand{\quater}{\mbox{\rm\hspace{1.2mm}\rule[0.05mm]{0.1mm}{2.5mm}\hspace{-1.2mm}Q}}

\newtheorem{prop}{Proposition} \renewcommand{\theequation}{\arabic{section}.\arabic{equation}}
\setcounter{equation}{0} \input epsf %--------------------------------------

\vskip 7cm

\centerline{{\Large Duality on Higher Order U(1) Bundles}}

\vskip 2cm

\centerline{{\large M. I. Caicedo, I. Mart\'{\i}n and A. Restuccia}}

\vskip .5cm

\centerline{\it Universidad Sim\'{o}n Bol\'{\i}var, Departamento de F\'{\i}sica}

\centerline{\it Apartado postal 89000, Caracas 1080-A, Venezuela.} \vskip .5cm

\centerline{\it e-mail: mcaicedo@usb.ve,isbeliam@usb.ve, arestu@usb.ve}

\vskip 2cm

{\bf Abstract}

\begin{quotation}

\noindent{\small A new global approach in the study of duality transformations
is introduced. The geometrical structure of complex line bundles is generalized to higher order U(1) bundles which
are classified by quantized charges and duality maps are formulated over these structures. Quantum
equivalence is shown between dual theories. A global constraint is proven to be needed to achieve
well defined bundles. These global structures are used to refine the proof of the duality equivalence
between d=11 supermembrane and d=10 IIA Dirichlet supermembrane, giving a complete
topological interpretation to their quantized charges.} 

\end{quotation}

\newpage

%\vskip 1.5cm

\section{Introduction}

The usual electromagnetic duality concept introduced first by Dirac in his dissertation on
magnetic monopoles, later extended by Montonen and Olive and used, lately, by Seiberg and Witten
{\cite{re:Witten}}to discuss the strong and weak coupling limits of the low energy effective
action of N=2 SUSY SU(2) Yang-Mills Theory , has provided a breakthrough in the understanding of
the non-perturbative analysis of QFT. Also, it has given a powerful tool to unify different
superstring and supermembrane theories  and to possibly  merge them in the context of M-theory, a
hypothetical theory of membranes and 5-branes whose low energy effective action is $d=11$
supergravity {\cite{re:Ferrara}}. Duality, in the above sense, may be understood as a map between
two quantum equivalent $U(1)$ gauge theories, one of them formulated in terms of a $U(1)$ 1-form
connection $A$ and coupling constant $\tau$ and its dual theory given by another $U(1)$ 1-form
connection $V$ and coupling constant $\frac{1}{\tau}$. The dual map being intrinsically
non-perturbative.

In the present article, we introduce the most general duality map between locally antisymmetric
fields, represented as local p-forms with non trivial transitions on intersections of open sets of
a covering of a compact d-dimensional manifold . We introduce the notion of higher order U(1)
bundles with a geometrical structure that generalizes  the U(1) principal bundles i.e. the usual
2-cocycle condition on the intersection of three open sets is elevated to a (p+1)-cocycle
condition on the intersection of  (p+2) open sets. The natural object generalizing a 1-form
connection on a U(1) bundle is a local p-form with non trivial transformation on the overlapping
of two open sets. Non trivial in the sense that it cannot be eliminated by a gauge transformation,
in the same way a gauge transformation cannot eliminate the non trivial transition on the
connection 1-form defining the magnetic monopole. We describe the  construction of higher order
bundles in section 3 and formulate the general dual map between dual actions. Furthermore we show
quantum equivalence of the two dual theories in section 4.

As a straightforward consequence of the above construction we obtain the generalized Dirac
quantization condition on the couplings. The non trivial higher order bundles naturally describe
quantized charges and are the appropriate geometrical objects to formulate the antisymmetric
fields involved in D-branes theories. The charges here have a topological origin. This is
developed in section 5.

To show the quantum equivalence between  dual electromagnetic theories, one starts from a theory
defined over the space of all connection 1-forms on all line bundles over the base 4-manifold $X$
and then go to an equivalent formulation in terms of globally defined constrained 2-forms. From
here one proceeds to introduce via Lagrange multipliers the dual connection 1-forms $V$ and after
functional integration, the dual theory is straightforwardly obtained. This approach may be
synthesized in the following sequence: \begin{eqnarray*} A \Leftrightarrow L_2 (constrained)
\Leftrightarrow V \end{eqnarray*} where $L_2 $ is the globally defined constrained 2-form. In
general, the dual maps on higher order bundles are defined through a similar sequence:
\begin{eqnarray*} A_p \Leftrightarrow L_{p+1} (constrained) \Leftrightarrow V_{d-p-2}
\end{eqnarray*} where $V_{d-p-2}$ represents the dual antisymmetric field. $L_{p+1}$ is
constrained to be a closed form but, in addition, it is globally restricted as well .This  global
constraint has to be implemented from the {\it {beginning}} in the mechanism to prove on shell
global equivalence and quantum equivalence between dual theories.

For dual maps between 1-form connections in four dimensions the global restriction is the usual
Dirac quantization condition, in other cases, as for the $d=11$ supermembrane $\Leftrightarrow$
$d=10$ Dirichlet supermembrane equivalence, the global constraint becomes the compactification
condition on one of the supermembrane coordinates. In general, the global condition contains the
relevant physical parameters involved in the duality map.

\setcounter{equation}{0} \section{Duality on The Space of Connections on Line Bundles}

Duality transformations among connections on  $U(1)$ line bundles over a manifold $X$ requires the
use of a  quantum equivalent formulation in terms of an independent 2-form. The original theory,
expressed as a functional in the space of abelian connections, is reformulated in terms of a
2-form constrained by  non-local and local conditions which ensure the existence of a
correspondence between the space of constrained 2-forms and the line bundles over the base space
$X$.

The purpose of this section is twofold, in the first place, we will construct the quantum
formulation of Maxwell's theory in terms of a {\it {globally}}  constrained 2-form and explicitly
show its equivalence to the usual connection formulation. In second place, using the above
formulation we will show  the duality between two $U(1)$ theories, one of them with coupling
constant $\tau$ and the other with coupling $-\frac{1}{\tau}$.( Some contents 
of this section were first obtained by {\cite{re:Mario}}).

We begin by considering Maxwell's theory, formulated in terms of a connection 1-form $A$ of a
$U(1)$ bundle $L$ with base space $X$ -a four dimensional compact orientable euclidean manifold-,
with these objects the theory is defined by the following action

\be
I(F(A))=\frac{1}{8\pi}\int_{X}d^4x\sqrt{g}[\frac{4\pi}{e^2}F^{mn}F_{mn}+i\frac{\theta}{4\pi}\frac{1}{2}\epsilon_{mnpq}F^{mn}F^{pq}]
\ee

Duality is usually addressed in terms of the action of the modular group $SL(2,\entero{})$ on the
complex coupling constant $\tau\equiv\frac{\theta}{2\pi}+i\frac{4\pi}{e^2}$. Upon introduction of
$\tau$ and using the standard decomposition of the curvature ($F=dA$) in its self-dual and anti
self-dual parts, $I(F(A))$ can be reexpressed as follows

\be I_\tau(A)=\frac{i}{8\pi}\int_X{}d^4x\sqrt{g}[\bar{\tau}F_{mn}^+F^{+mn}
-\tau{}F_{mn}^-F^{-mn}]\ , \ee

or in terms of the inner product of forms

\be I_\tau(A)=\frac{i}{4\pi}[\bar{\tau}(F^+,F^+)-\tau{}(F^-,F^-)] \label{eq:itau} \ee

We now introduce an action where the independent field is an arbitrary 2-form $\Omega$, globally
defined over $X$ as follows

\be I(\Omega{})=\frac{i}{4\pi}[\bar{\tau}(\Omega^+,\Omega^+)-\tau{}(\Omega^-,\Omega^-)]
\label{eq:iomeg} \ee

The quantum field theories associated to actions (2.3) and (2.4) are not equivalent since $\Omega$
is arbitrary i.e. $I(\Omega{})$ is a functional over the whole space of 2-forms while $F$ is the
curvature of a $U(1)$ connection.

As a first step in our programme, we will show that after restricting the space of 2-forms in
(2.4) by the introduction of two constraints, the theories defined by (2.3) and the constrained
version of (2.4) are equivalent as QFTs. The constraints to be imposed on $\Omega$ are

\bea d\Omega{}=0\label{eq:locvinc}\\ \oint_{\Sigma_2}\Omega=2\pi n\label{eq:globvinc} \eea

where $\Sigma_2$ represents a basis of the integer homology of dimension 2 over $X$, to each
element of the basis we associate an integer $n$.

The first of these constraints restricts $\Omega$ to be closed, while the second ensures its
periods to be integers (Dirac's quantization condition).

The second step in the discussion is to show that if one introduces a new line bundle -to which we
will refer to as  the dual line bundle $^*\!{}L$- with connection 1-form $V$, it is possible to
include constraints (2.5) and (2.6) in the action through the appropriate use of $V$ as a Lagrange
multiplier in the following fashion

\be {\cal{I}}(\Omega,V)=I(\Omega)+\frac{i}{2\pi}\int_{X}W(V)\wedge\Omega \label{eq:ivinc} \ee

\noindent{}where, $W(V)\equiv{}dV$ is the curvature associated to $V$. Before we engage in the
rigorous proof of the above claims, we note that -as we shall see later on- had one used the usual
constraint $\int_XV\wedge{}d\Omega$, the constraint in the periods of $\Omega$, which is a global
condition, would not have been obtained.

We begin by considering constraints (2.5) and  (2.6). If $F(A)$ is the curvature associated to a
connection 1-form then it is obviously closed, i.e. satisfies the local constraint (2.5);
moreover, the requirement on the transition functions of the line bundle to be uniform maps over
the structure group guarantees that $F(A)$ also satisfies the global constraint (2.6). The
following proposition, shows that the converse is also true:

{\it{If $\Omega$ is a 2-form satisfying constraints (2.5) and (2.6) then there exists a  complex
line bundle and a connection -not necessarily unique- on it whose curvature is $\Omega$}}
\cite{re:weil}

\noindent{} Let ${\cal{U}}=\{U_i/i\in{}I\}$ be a contractible covering of $X$. The condition of
closeness on $\Omega$ guarantees that in $U_i\bigcap{}U_j\bigcap{}U_k\neq\emptyset$ , $\Omega$ may
be written as:

\be \Omega=dA_i=dA_j=dA_k \ee

which implies

\bea A_i=A_j+d\lambda_{ij}\nonumber\\ A_j=A_k+d\lambda_{jk}\\ A_k=A_i+d\lambda_{ki}\nonumber \eea

From here we get

\be \lambda_{ij}+\lambda_{jk}+\lambda_{ki}=constant=c \ee

The global condition on the periods of $\Omega$ leads to, see (3.14) to (3.21) for details,

\be c=2\pi n \ee

We thus conclude that in the sense of \v{C}ech the 2-cochain \cite{re:Steenrod} \be
g:(i,j)\rightarrow{}g_{ij}\equiv{}e^{i\lambda_{ij}} \in{}U(1)\ee

is a 2-cocycle

\be \delta{}g_{ijk}=g_{ij}g_{jk}g_{ki}=\Uno \ee

Moreover, if one changes $A$ by a gauge transformation

\bea &A_i\rightarrow{}A_i+d\lambda_i\hbox{ in }U_i&\nonumber\\ &\hbox{}&\\
&A_j\rightarrow{}A_j+d\lambda_j\hbox{ in }U_j&\nonumber \eea

then  \be \lambda_{ij}\rightarrow{}\lambda_{ij}+\lambda_i-\lambda_j \label{eq:delta1} \ee

therefore  $g_{ij}$ changes as

\be g_{ij}\rightarrow{}h_ig_{ij}h^{-1}_j \ee

\noindent{}Now we notice that $h_ih_j^{-1}$ is a coboundary as follows from the fact that

\be h:(i)\rightarrow{}h_i=e^{i\lambda_i}\in{}U(1) \ee

\noindent{}is a map from $U_i$ to the structure group, and

\be \delta{}h(i,j)=h_ih_j^{-1} \ee

\noindent{}consequently, under (2.15) $g_{ij}$ changes by a coboundary, and then it defines the
same element of the \v{C}ech cohomology $H^1(\cal{U} , C*)$. $C*$ is the set of
non zero complex numbers.
It is known \cite{re:Steenrod} that there is a one-to-one correspondence
between $H^1(\cal{U}, C*)$ and the complex line bundles over $X$, $g_{ij}$
defining the transition functions of the bundle. Constraints (2.5) and (2.6) define then an unique line bundle over $X$.
Moreover $A$, defined by patching together the 1-forms $A_{i}$ on
$U_{i}$ by using (2.14), is a connection 1-form over $X$ and $\Omega$ its curvature 2-form.

Regarding the non uniqueness of the connections on the line bundle associated
to $\Omega$, one must realize that two connection 1-forms $A_1$ and $A_2$  with the same curvature
may be in different equivalence classes not related by gauge
transformations.They differ at most by a closed 1-form $\theta \in{}H^1(X,R)$.
If $\theta$ is an element of $H^1(X,Z)$ then $A_1$ and $A_2$ are connections on
the same equivalence class but otherwise they belong to different ones.
 The equivalence classes of connections  related to the same  $\Omega$ are in one-to-one  
 correspondance  to  $ H^1(X,U(1))$. Moreover, one
has for the holonomy maps $Q$ constructed with connections with the same curvature $\Omega$,
\begin{eqnarray*} Q^{\chi .l} = \chi Q^l \end{eqnarray*} here $l$
denotes a line bundle with a particular equivalence class of connections and
$\chi$ is the holonomy map given by the exponential of the integral of $\theta $  around a closed curve. 
For a simple connected base manifold $X$ the  line bundle
associated to $\Omega$ is unique. {\cite {re:weil}} The observation just made is relevant to the
proof of the quantum equivalence of the theories defined by $I_\tau{}(A)$ and $I(\Omega)$
restricted by the constraints we have been studying. Indeed, when formulating the quantum
correlation functions for either theory, one must carefully define the functional measure in order
to account for the "zero modes", that is the space $ H^1(X,R)$.

It is worth noticing  that -up to the definition of the measure-, the equivalence of the quantum
theories rests on the non local constraint on the periods of the 2-form $\Omega$. There is no
local formulation of Maxwells theory ($I_\tau(A)$) in terms of $\Omega$. The local restriction
$d\Omega{}=0$ is not sufficient to guarantee the existence of a  line bundle and a connection with
curvature $\Omega$. The global constraint associates a set of integers $\{n\}$ (the winding
numbers or topological charges) to the elements of a basis of homology of dimension 2.

In order to continue with the proof of the quantum equivalence, we come to study the formulation
of the off shell Lagrange problem  associated to (2.4), (2.5) and (2.6). We will see that (2.4),
subject to (2.5) and (2.6), and (2.7) are equivalent when summation over all line bundles is
considered in the functional integral.

We first consider the extra piece in ${\cal{I}}(\Omega{},V)$ i.e.

\be {\cal{S}}_{Lagrange}=\frac{i}{2\pi}\int_XdV \wedge \Omega{} \ee

Where we must recall that $V$ is a connection 1-form on the dual bundle $L^*$.
${\cal{S}}_{Lagrange}$ can be rewritten as

\be {\cal{S}}_{Lagrange}=-\frac{i}{2\pi}\int_X{}V \wedge
d\Omega{}+\frac{i}{2\pi}\int_Xd(V\wedge\Omega{}) \label{eq:extra} \ee

The functional integration on $V$ may be performed in two steps. We first integrate on all
connections over a given complex line bundle and then over all complex line bundles. The second
term on (2.20) depends only on the transition function of a given complex line bundle, while the
first depends also on the space of connections over the line bundle. Integration associated to the
first step yields a

\be \delta{}(d\Omega{}) \ee on the functional measure.

It is convenient to rewrite the second term in (2.20) as

\be \frac{i}{2\pi}\int_Xd(V\wedge\Omega{})
=\frac{i}{2\pi}\int_{\Sigma_{3}}(V_+-V_-)\wedge\Omega{}=
\frac{i}{2\pi}\int_{\Sigma_3}d\xi_{+-}\wedge\Omega{}=i n \int_{\Sigma_{2}}\Omega \ee

where $\Sigma_3$ denotes 3-dimensional surfaces living in the intersection of open sets where the
transition of the connection 1-form $V$ takes place,

\bea &V_+-V_-=d\xi_{+-}&\nonumber\\ &g_{+-}=e^{i\xi_{+-}}&\nonumber \eea

$g_{+-}$ being the transition function and $\xi_{+-}$ is, in general, a multivalued function.

Summation over all line bundles gives from (2.22), and after Fourier transforming,

\be \sum_m \delta{}(\int_{\Sigma_{2}}\Omega{}-2\pi{}m) \ee

where $\Sigma_{2}$ denotes a basis of an integer homology of dimension 2. We thus conclude that
the Lagrange problem associated to (2.4),(2.5) and (2.6) is given by the action (2.7).

We turn now to the discussion of the full partition function associated to the actions
$I(F(A))$,$I(\Omega{})$ subject to (2.5) and (2.6), and ${\cal{I}}(\Omega,V)$. The path integral
that defines the problem is given by

\be Z_1=\sum_
\int{\cal{D}}\Omega{\cal{D}}V{\hbox{Vol}}_{ZM}{\hbox{det}}(d_2)\frac{1}{{\hbox{Vol}}G}e^{-{\cal{I}}(\Omega,V)}
\ee

\noindent{}where as we have just learned , $\sum_n$ stands for summation over all line bundles.
${\hbox{Vol}}_{ZM}$ is the volume of the space $ H^1(X,R)$, det($d_2$) is the determinant of the
exterior differential operator on 2-forms and Vol $G$ is the volume of the gauge group. After
performing the integration on $V$ as described we obtain

\be {\cal{Z}}(\tau{})=\sum_m
\int{\cal{D}}\Omega{\hbox{Vol}}_{ZM}{\hbox{det}}(d_2)\delta{}(d\Omega{})\delta{}(\int_{\Sigma_{2}}
\Omega{}-2\pi{}m)e^{-I(\Omega)} \ee

The measure may now be reexpressed in terms of an integration on the space of connections $A$ over
the line bundle $L$ in the following way

\be
{\cal{Z}}(\tau{})=\sum_m\int{\cal{D}}\Omega{\hbox{Vol}}_{ZM}\int{\cal{D}}A\frac{1}{{\hbox{Vol}}G}
\frac{\delta{}(\Omega{}-F(A))}{{\hbox{Vol}}_{ZM}} \delta{}(\int_{\Sigma_{2}}\Omega{}-2\pi{}m)
e^{-I(\Omega)} \ee

The factor $1/{\hbox{Vol}}_{ZM}$ that comes from reexpressing $\delta{}(d\Omega{})$ in terms of
$\delta{}(\Omega{}-F(A))$ exactly cancels the volume originally appearing in the functional
measure. Further integration in $\Omega$  produces the final result

\be {\cal{Z}}(\tau{})={\cal{N}}\int{\hat{\cal{D}}}A\frac{1}{{\hbox{Vol}}G}e^{-I_\tau{}(A)}
\label{eq:YESS} \ee

Where $\hat{\cal{D}}A$ denotes integration over the space of connections on all line bundles over
$X$. Since (2.27) is the partition function for the action $I_\tau{}(A)$, we have been able to
show the quantum equivalence of the three formulations of Maxwell's theory thus finishing the
first part of our programme.

Finally, we would like to briefly discuss the duality transformations in the functional integral
associated to Maxwell's theory. We start from the action

\be {\cal{I}}(\Omega,V)=\frac{i}{4\pi}[\bar{\tau}(\Omega^+,\Omega^+)-\tau{}(\Omega^-,\Omega^-)]
+\frac{i}{2\pi}(W^+(V),\Omega^+)\frac{i}{2\pi}(W^-(V),\Omega^-), \ee

\noindent{}from where it is possible to perform the functional integration on $\Omega^+$ and
$\Omega^-$ to get the known result \cite{re:Witten2}

\be {\cal{Z}}(\tau{})={\cal N}
\tau^{-\frac{1}{2}B^-_2}\bar{\tau}^{-\frac{1}{2}B^+_2}{\cal{Z}}(-\frac{1}{\tau}) \label{eq:dualz}
\ee

\noindent{}where $B^+_k$ and $B^-_k$ are the dimensions of the spaces of selfdual and antiselfdual
$k$ forms, this last formula can be reexpressed in terms of the Euler characteristic $\chi$ and
the Hirzebruch signature $\sigma$ as

\be [Im(\tau{})^{\frac{1}{2}(B_0-B_1)}{\cal{Z}}(\tau{})]= {\cal N}
\tau^{-\frac{1}{4}(\chi{}-\sigma{})}\bar{\tau}^{-\frac{1}{4}(\chi{}+\sigma{})}
[Im(-\frac{1}{\tau})^{\frac{1}{2}(B_0-B_1)}{\cal{Z}}(-\frac{1}{\tau}) \label{eq:dualz} \ee ${\cal
N}$ is a factor independent of $\tau$ that depends on the topology of $X$.

We have thus been able to implement the duality transformations in a rigorous way by including the
global constraint and the associated measure factors in the functional integral of the Maxwell
action over a general base manifold $X$.

\setcounter{equation}{0} \section{ Higher order U(1) bundles }

In the previous section we proved the existence of a line bundle associated to a closed, integer
2-form $L_2$ globally defined over $X$,in this section we present an extension of it.  We will
consider a closed integer p-form $L_p$ globally defined over $X$ -an orientable compact euclidean
manifold- and show the existence of an associated geometrical structure characterized by
(p-1)-forms with values in the $U(1)$ algebra and transitions (p-2)-forms satisfying the cocycle
condition on the intersection of (p+1) open neighborhoods of a covering of $X$.

We start with the case $p=1$, this is relevant to show the equivalence between the $d=11$
supermembrane and the $d=10$ IIA Dirichlet supermembrane as we will discuss in the next section,
and then go to $p\geq3$ cases.

Let $L_1$ be a 1-form globally defined over $X$ satisfying

\begin{eqnarray} dL_{1} & = & 0 \\ \oint_{\Sigma_{1}}L_1 & = & 2\pi n \end{eqnarray} where
$\Sigma_1$ is a basis of an integer homology of curves over $X$ and  $n$ is an integer associated
to each element of the basis, then $L_1$ must satisfy the following equation \begin{equation}
L_{1}=-ig^{-1}dg \end{equation} where \begin{equation} g=\exp{i\varphi} \end{equation} defines an
uniform map from \begin{equation} X\rightarrow S^{1}¥ \end{equation} $\varphi$ being an angular
coordinate on $S^1$.

Conversely, given $g$ an uniform map from $X\rightarrow S^1$ then (3.3) defines a closed 1-form
with integer periods. Let us prove the above claim, given $L_1$ globally defined over $X$
satisfying (3.1) and (3.2) we may define

\begin{equation} \varphi(P)=\int_{O}^{P}L_{1} \end{equation}

where $O$ and $P$ are the two end points of a curve on the base manifold $X$, $O$ being a
reference point. $\varphi(P)$ is independent of the curve within a homology class in the sense
that

\begin{eqnarray*} \int_{{\cal C}} L_{1}=\int_{{\cal C }^\prime} L_{1} \end{eqnarray*} if the
closed curve ${\cal C}^{-1}.{\cal C}^{\prime}$ is homologous to zero and, by assumption (3.2),
differs in $2\pi n$ between two different homology classes.  ${\cal C}$ and ${{\cal C }^\prime}$
are open curves with the same end points. (3.4) then defines an uniform map from $X\rightarrow
S^1$. The converse follow directly by the same arguments.

We may also understand (3.6) from a different point of view by considering a covering of $X$ with
open sets $U_i$, $i\epsilon I$.  We may always assume $U_i$ and $U_i\cap U_j \neq \emptyset$, $i,j
\epsilon I$ to be contractible to a point.

On $U_i$, $i\epsilon I$ \begin{equation} L_{1}=d\lambda^{i}, \end{equation} and on $U_i\cap U_j
\neq \emptyset$

\begin{eqnarray} d\lambda^{i} & = & d\lambda^{j}\\ \lambda^{i} & = &\lambda^{j}+c^{ji}
\end{eqnarray} where $c^{ji}$ is a constant on the intersection.

We may define on $U_j$

\begin{equation} \lambda^{\prime j} = \lambda^{j}+c^{ji} \end{equation} without changing $L_1$ and
with a trivial transition on $U_i\cap U_j \neq \emptyset$.

We may extend $\lambda^i$ to $U_j$, and so on from $U_j$ to $U_k$, $U_j\cap U_k \neq \emptyset$,
until we meet an $U_k$ such that  \begin{eqnarray*} U_{k}\cap U_{i}\neq \emptyset
\end{eqnarray*}where we cannot redefine $\lambda^i$.  If we denote

\begin{equation} L_{1}=d\lambda \end{equation} we arrive to a multivalued function $\lambda$.
Condition (3.2) ensures that the transition (which in this case defines the multivaluedness of
$\lambda$) is $2\pi n$. We thus obtain (3.3), (3.4) and (3.5).

Let us consider again the case of a 2-form $L_2$ already discussed in section 2, we would like to
add some remarks on it.  Let $L_2$ be a 2-form on $X$ satisfying \begin{eqnarray} dL_{2} & =& 0\\
\oint_{\Sigma_{2}}L_{2} &=& 2\pi n \end{eqnarray} where $\Sigma_2$ denotes a
basis of an integer homology of dimension 2. Then there exists a complex line bundle over $X$ and

\begin{equation} L_{2}=dA \end{equation} is the curvature of a connection 1-form A
on open sets of a covering. Conversely,
given a line bundle over $X$ and a connection 1-form A its curvature satisfies (3.12) and (3.13).

On $U_i$, $i\epsilon I$,we have

\begin{equation} L_{2}=dA_{i} \end{equation} similarly, on  $U_i\cap U_j \neq \emptyset$

\begin{equation} A_{i}=A_{j}+d\omega_{ij}¥ \end{equation} and on $U_i\cap U_j\cap U_k \neq
\emptyset$ \begin{equation} \omega_{ij}+\omega_{jk}+\omega_{ki}= constant. \end{equation}

Hence if we take a $\Sigma_2$ in the intersection $U_i$, $U_j$ and $U_k$, see
Figure 1, \begin{figure}[h] \epsfxsize=2in \centerline{\epsffile{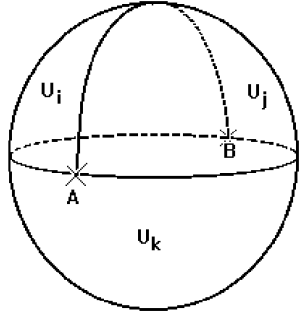}} \caption {Intersecting open sets} \end{figure}
we obtain \begin{equation} \int_{\Sigma_2}L_{2}=\int_{\Sigma_{1}}L_{1}=\displaystyle \left.
(\omega_{ij}+\omega_{jk}+\omega_{ki}) \right|_{B}^{A}	=2\pi n \end{equation} where $\Sigma_1$ is
the union of the three curves on the figure and $L_1= d (\omega_{ij} +\omega_{jk} +\omega_{ki})$.
Without loosing generality we may redefine the $\omega$ such that the value of the parenthesis in
(3.18) at B is zero.  We thus obtain the cocycle condition \begin{equation}
\omega_{ij}+\omega_{jk}+\omega_{ki}=2\pi n	. \end{equation}

We thus have associated to each  $U_i\cap U_j \neq \emptyset$ a map $g_{ij}= \exp{(i\omega_{ij})}$

\begin{equation} U_{i}\cap U_{j}\rightarrow S^{1}, \end{equation} satisfying the cocycle condition
on $U_i\cap U_j\cap U_k \neq \emptyset$:

\begin{equation} g_{ij}g_{jk}g_{ki}=\Uno . \end{equation}

Let us now consider a 3-form $L_3$ globally defined on $X$ satisfying

\begin{eqnarray} dL_{3} & = & 0 \\ \oint_{\Sigma_{3}}L_3 &=& 2\pi n. \end{eqnarray} where
$\Sigma_{3}$ is a basis of an integer homology of dimension 3. As in previous cases to each
element of the basis one associates an integer $n$.

We now have on $U_i$ \begin{equation} L_{3}=dB_{j} \end{equation} and on $U_i\cap U_j \neq
\emptyset$ \begin{eqnarray} dB_{i} &= &dB_{j}  \nonumber\\ B_{i}& = & B_{j}+ d\eta_{ij}
\end{eqnarray} where $B_i$ is a 2-form with transition given by (3.25), $\eta_{ij}$ being a local
1-form defined on $U_i\cap U_j$. On $U_i\cap U_j\cap U_k$ we obtain \begin{eqnarray} L_{1} &
\equiv & \eta_{ij}+\eta_{jk}+\eta_{ki}\nonumber \\ d L_{1}&  = &0 \end{eqnarray}

In order to determine the periods of $L_1$ we proceed as in (3.17)-(3.19) where we determined the
value of the constant for the 0-form  $\omega_{ij} + \omega_{jk} + \omega_{ki}$.We consider a
$\Sigma_3$ intersecting $U_i$, $U_j$ and $U_k$.

We then have from (3.23) \begin{equation} \int_{\Sigma_{3}}L_{3}=\int_{\Sigma_{1}}L_{1}=2\pi n,
\end{equation} where $\Sigma_{1}$ is a closed curve on $U_i\cap U_j\cap U_k$.  From (3.26) and
(3.27), we thus obtain a 1-form $L_1$ defined over $U_i\cap U_j\cap U_k$ satisfying (3.1) and
(3.2) which yields an uniform map $M$ from \begin{equation} U_i\cap U_j\cap U_k \rightarrow U(1)
\end{equation}

The interesting property not present in the previous discussion is that the 1-cochain is now
defined as \be g:(i,j) \rightarrow g_{ij}(P,{\cal C}) \equiv \exp i\int_{\cal C}\eta_{ij} \ee
where $\cal C$ is an open curve with end points $O$ (a reference point) and $P$. $g$ associates to
$(i,j)$ a map $g_{ij}(P,\cal C)$ from the path space over $U_i\cap U_j$ to the structure group
U(1).

Notice that the 1-form $\eta_{ij}$ cannot be integrated out to obtain a transition function as in
the case of a line bundle. However, we have \be \delta{}g_{ijk}=g_{ij}g_{jk}g_{ki}=\exp
i\int_{O}^{P}L_{1} \ee which is precisely the uniform map $M$ previously defined in (3.28). (3.29)
explicitly shows that the geometrical structure we are dealing with is not that of an usual $U(1)$
bundle since the cocycle condition on the intersection of three open sets of the covering is not
satisfied. Starting from transitions functions $g_{ij}$ defined on the space of paths over
$U_i\cap U_j$, and acting with the coboundary operator $\delta$ we obtain the 2-cochain (3.30)
which is properly defined in the sense of \v{C}ech. We may go further and consider in the
intersection of four open sets the action of the coboundary operator $\delta$ on 2-cochains. We
wish to construct now a 3-cocycle on that intersection. We have from (3.26) on $U_i\cap U_j\cap
U_k\cap U_l$

\begin{eqnarray} d\lambda_{ijk} & = & \eta_{ij}+\eta_{jk}+\eta_{ki} \nonumber \\ d\lambda_{ijl} &
= & \eta_{ij}+\eta_{jl}+\eta_{li}\nonumber \\ d\lambda_{ikl} & = &
\eta_{ik}+\eta_{kl}+\eta_{li}\nonumber \\ d\lambda_{jkl} & = &
\eta_{jk}+\eta_{kl}+\eta_{lj}\nonumber\end{eqnarray}

which implies

\begin{equation} d( \lambda_{ijk} - \lambda_{ijl} + \lambda_{ikl} - \lambda_{jkl})  =  0 \nonumber
\end{equation}

Using (3.27) we finally obtain,

\begin{equation}  \lambda_{ijk} - \lambda_{ijl} + \lambda_{ikl} - \lambda_{jkl}  = 2\pi n
\nonumber \end{equation}

We then define the 2-cochain on $U_i\cap U_j\cap U_k$

\begin{equation} g:(i,j,k)\rightarrow g_{ijk}\equiv\exp{i\lambda_{ijk}} \end{equation}

it satisfies the 3-cocycle condition

\be \delta{}g_{ijkl}=g_{ijk}g^{-1}_{ijl}g_{ikl}g^{-1}_{jkl}=\Uno \ee

The geometrical structure we are introducing is defined by equations (3.29),(3.30) and (3.34). It
generalizes the geometrical structure on a principal bundle and it is the natural one to consider
in the context of dual maps. In particular it is the geometrical structure associated to the
D-brane actions as we will discuss later.

The procedure may be generalized to globally defined p-forms $L_p$ over $X$, satisfying
\begin{eqnarray} dL_{p} & =& 0 \nonumber\\ \oint_{\Sigma_{p}}L_{p} &=& 2\pi n. \end{eqnarray} This
gives a geometrical structure with transition p-2 forms $\eta$ with values on the Lie algebra of
the structure group leading to 1-cochains

\begin{equation} \exp{i\int_{\Sigma_{p-2}}\eta} \end{equation} $L_p$ being the curvature of a
local p-1 form with transitions given by $d\eta$.  Moreover on $U_i\cap U_j\cap U_k \neq
\emptyset$ the p-2 transition form \begin{equation} L_{p-2}=\eta_{ij}+\eta_{jk}+\eta_{ki}
\end{equation} satisfies the conditions \begin{eqnarray} dL_{p-2} & =& 0 \nonumber \\
\oint_{\Sigma_{p-2}}L_{p-2} &=& 2\pi n \end{eqnarray} and hence the structure of $L_{p-2}$ may be
determined by induction. We end up with a p-cocycle condition on the intersection of $p+1$ open
sets. Summing up, we have shown the existence of local antisymmetric fields with non trivial
transition conditions generalizing the structure of connection 1-forms over complex line bundles.

In section 5, we will apply these results to show quantum equivalence of the $d=11$ supermembrane
and the $d=10$ IIA Dirichlet supermembrane for the general case of non trivial line bundles
associated to the U(1) gauge fields in the Dirichlet supermembrane multiplet.  We will thus extend
previous proofs valid for trivial line bundles.

\setcounter{equation}{0} \section{ Duality in higher order $U(1)$ bundles}

In this section, we discuss the general duality map relating local antisymmetric fields defined
over higher order $U(1)$ bundles. Duality with p-forms on trivial bundles was
first analysed by Barb\'on \cite{re:Witten2}. The action for the local $U(1)$ p-form $A_p$ defined over open sets of a covering of $X$, a compact manifold of dimension $d\ge p+1$ with $p >1$, and with
transitions given as in section 3, is the following \be {S(A_p)}= \frac{1}{2}\int_X{}F_{p+1}\wedge
*F_{p+1}{}+ g_p \int_{\Sigma_{p}}  A_p{} \ee where $F_{p+1}$ is the globally defined curvature
(p+1)-form associated to $A_p$. $\Sigma_p$ is a p-dimensional closed surface being the boundary of
a (p+1)-chain. $g_p$ is the coupling associated to $A_p$. From (4.1) we obtain the field equations
\be d*F_{p+1} = g_p\delta {\Sigma_p} \ee where $\delta {\Sigma_p}$ is the usual (d-p)-form
associated to the Dirac density distribution.

Let us consider now the dual formulation to (4.1). Following the arguments of the previous
sections, we introduce a constrained (p+1)-form $L_{p+1}$ globally defined over $X$ satisfying
\begin{eqnarray} dL_{p} & =& 0 \nonumber\\ \oint_{\Sigma_{p}}L_{p} &=& \frac{2\pi n}{g_p}.
\end{eqnarray} with action \be {S}= \frac{1}{2}\int_X{}L_{p+1}\wedge *L_{p+1}{}+ g_p
\int_{\Sigma_{p+1}} L_{p+1} \ee where $\Sigma_{p+1}$ is a (p+1)-chain with boundary $\Sigma_{p}$.

The off-shell Lagrange problem  of the above constrained system may be given by the action \be
S(L_{p+1},V_{d-p-2}) = S(L_{p+1})+ i \int_X{}L_{p+1}\wedge W_{d-p-1}(V){} \ee where $W_{d-p-1}$ is
the curvature of the local (d-p-2)-form $V_{d-p-2}$ defined over a higher order bundle satisfying
the (d-p-1)- cocycle condition introduced in section 3. Consequently, $W_{d-p-1}$ satisfies
identically the conditions
 \begin{eqnarray} dW_{d-p-1} & = & 0 \nonumber\\ \oint {W_{d-p-1}} &=&
\frac{2\pi n}{g_{d-p-2}}. \end{eqnarray}
 Integration on $V_{d-p-2}$ leads to the action (4.1)
while integration on $L_{p+1}$ yields the on-shell condition \be *L_{p+1} = - i W_{d-p-1} - g_p
\delta {\Sigma_{p+1}} \ee and the dual action \be S(V_{d-p-2}) =
\frac{1}{2}\int_X{}W_{d-p-1}(V)\wedge *W_{d-p-1}{} + g_{d-p-2}\int_{\Sigma_{d-p-2}} V_{d-p-2} \ee
where \begin{eqnarray*}\int_{\Sigma_{d-p-2}}\cdot=-\frac {g_p}{g_{d-p-2}}\int_{X}
d(*\delta(\Sigma_{p+1}))\cdot  \end{eqnarray*} From (4.2) and (4.7)  we obtain the quantization
condition \be g_p g_{d-p-2} = 2\pi n \ee

The quantum equivalence of the dual actions (4.1) and (4.8) follows once one integrates over all
corresponding higher order bundles. This is a generalization of the equivalence proven in section
2 for the electromagnetic duality. The quantization of charges is
directly related to the different higher order bundles that may be constructed over $X$ and it
arises naturally from the global constraint (4.3) needed for having a globally well defined bundle.
The correspondance between closed integral p-forms and bundles is in general not one-to-one,
depending on the topology of the base manifold, the redundancy being given by
$H^{p-1}(X, U(1))$.

\setcounter{equation}{0} \section{ Global analysis of duality maps in p-brane theories}

We use in this section the  global arguments of the previous sections to improve the p-brane
$\Leftrightarrow$ d-brane equivalence that has been proposed by {\cite{re:Town}\cite{re:
x6}\cite{re: x7}}. The duality transformation has been recently used by Townsend \cite{re:Town} to
show the equivalence between the covariant $d=11$ supermembrane action with one coordinate
$X^{11}$ compactified on  $S^1$, and the fully $d=10$ Lorentz covariant worldvolumen action for
the $d=10$ IIA Dirichlet supermembrane.The equivalence between the bosonic sectors was previously
shown by Schmidhuber \cite{re: x7} using the Born-Infeld type action found by Leigh \cite{re: x6}.
We will  argue in a global way showing the equivalence between both theories, even when nontrivial
line bundles are included in the  construction of the D-brane action. We discuss later on the
equivalence of the bosonic sectors when the coupling to background fields is included. We consider
following  \cite{re:Town} the Howe-Tucker formulation of the $d=11$ supermembrane over a target
manifold with one coordinate compactified on $S^1$ \cite{re: Howe}, that is we take $X^{11}$ to be
the angular coordinate $\varphi$ on $S^1$. The action is then

\begin{eqnarray} S&=&-\frac{1}{2}\int_X d^{3}\xi
\sqrt{-\gamma}[\gamma^{ij}\pi_{i}^{m}\pi_{j}^{n}\eta_{mn}+\gamma^{ij} (\partial_{i}\varphi
-i\bar{\theta}¥\Gamma_{11}\partial _{i}\theta)(\partial_{j}\varphi -i\bar{\theta}
\Gamma_{11}\partial _{j}\theta)-1] \nonumber \\ &-&\frac{1}{6}\int_X
d^{3}\xi\epsilon^{ijk}[b_{ijk}+3b_{ij}\partial_{k}\varphi] \end{eqnarray} where $\eta$ is the
Minkoswski metric in $d=10$ spacetime, and \begin{eqnarray} \pi^{m} & = &
dx^{m}-i\bar{\theta}\Gamma^{m}d\theta \nonumber\\ \epsilon^{ijk}b_{ijk}& = & 3\epsilon^{ijk}\{
i\bar{\theta}\Gamma_{mn} \partial_{i}\theta[\pi_{i}^{m}\pi_{j}^{n}+i
\pi_{i}^{m}(\bar{\theta}\Gamma^{n}
\partial_{j}\theta)-\frac{1}{3}(\bar{\theta}\Gamma^{m}\partial_{i}\theta)(\bar{\theta}\Gamma^{n}
\partial_{j}\theta)] \nonumber\\
&+&(\bar{\theta}\Gamma_{11}\Gamma_{m}\partial_{i}\theta)(\bar{\theta}\Gamma_{11}
\partial_{j}\theta)(\partial_{k}x^{m}-\frac{2i}{3}\bar{\theta}\Gamma^{m}\partial_{k} \theta)\}¥
\nonumber\\ \epsilon^{ijk}b_{ij}&=&-2
i\epsilon^{ijk}\bar{\theta}\Gamma_{m}\Gamma_{11}\partial_{i}\theta
(\partial_{j}x^{m}-\frac{i}{2}\bar{\theta}\Gamma^{m}\partial_{j}\theta). \end{eqnarray}

We will now perform the same steps as in section 2, 3 and 4. From section 3, the constraints
\begin{eqnarray} dL_{1} & = & 0 \\ \oint_{\Sigma_{1}}L_1 & = & 2\pi n, \end{eqnarray}

define a uniform map: $X\rightarrow S^1$

\begin{equation} g=\exp{i\varphi} \end{equation} and \begin{equation} L_1=-ig^{-1}dg=d\varphi.
\end{equation} The converse being also valid. In this context $\Sigma_1$ is a basis of homology on
the $d=3$ worldsheet manifold.

The intermediate step in the construction of the duality map consists then in attaining an
equivalent formulation to (5.1) in terms of the global 1-form $L_1$. The important point now is to
realize that the Lagrange formulation of the constraints (5.3) and (5.4) may be obtained in terms
of a connection 1-form over the space of all non trivial line bundles, exactly as in section 2.

So we start with action \begin{eqnarray} S_{1} & = & -\frac{1}{2}\int_{X} d^{3}\xi
\sqrt{-\gamma}[\gamma^{ij}\pi_{i}^{m}\pi_{j}^{n}
\eta_{mn}+\gamma^{ij}(L_{1i}-i\bar{\theta}\Gamma_{11}\partial _{i}\theta) (L_{1j}
-i\bar{\theta}\Gamma_{11}\partial _{j}\theta)-1] \nonumber\\ &- & \frac{1}{6}\int_{X}
d^{3}\xi\epsilon^{ijk}[b_{ijk}+3b_{ij}L_{1k}]+\int_{X} F(A)L_{1} \end{eqnarray} where $L_1$ is a
globally defined 1-form over $X$ and $A$ is a connection on the space of all line bundles over
$X$.

Functional integration on $A$ yields, by a similar argument to the one used in the analogous
problem we discussed in section 2, \begin{equation} \delta(dL_{1})\delta(\oint_{\Sigma_{1}}L_1 -
2\pi n) \end{equation} in the functional measure of the path integral.

We now use \begin{equation} \delta(dL_{1})\delta(\oint_{\Sigma_{1}}L_1 - 2\pi n)=\int [d\varphi]
\frac{\delta (L_1 -d\varphi )}{\det d} \end{equation} where $\varphi$ defines a map from
$X\rightarrow S^1$, that is $d\varphi$ satisfies (5.4). We notice that the functional integral in
(5.9) is over all maps from $X\rightarrow S^1$, it is not an integration over a cohomology class
defined by an element of $H^1(X)$.

In distinction to section 2,  zero modes, in this case, are constants. We may hence directly
integrate on $L_1$ and replace in (5.7) $L_1$ by $d\varphi$. We thus arrive to the covariant
$d=11$ supermembrane action after elimination of $L_1$.

On the other hand, we may functionally integrate $L_1$ in (5.7) to arrive to the functional
integral of the action \begin{eqnarray} S_{2}& =& -\frac{1}{2}\int_{X} d^{3}\xi
\sqrt{-\gamma}[\gamma^{ij}\pi_{i}^{m}\pi_{j}^{n} \eta_{mn}-\gamma^{ij}f_i(A) f_j(A) -1]
\nonumber\\ & - & \frac{1}{6}\int_{X} d^{3}\xi\epsilon^{ijk}b_{ijk}+\int_{X}d^{3}\xi
\gamma^{ij}f_i(A) i\bar{\theta} \Gamma_{11}\partial _{l}\theta \end{eqnarray} Where
\begin{eqnarray} f_i(A) \equiv \epsilon_{imn}(F^{mn}(A)-\frac{1}{2}b^{mn}). \end{eqnarray}

The functional integral in $A$ must now be performed over all line bundles over $X$. The result
(5.10) was obtained by Townsend in {\cite{re:Town}} , for the case of a trivial line bundle.  The
equivalence between (5.10), the fully $d=10$ Lorentz covariant worldvolume action for the $d=10$
IIA Dirichlet supermembrane, and the $d=11$ covariant supermembrane action (5.1) has then been
established.  In the functional integral for (5.1), integration over all maps between
$X\rightarrow S^1$ must be performed while in the functional integral for (5.10) integration over
the space of all connection 1-forms on all line bundles (modulo gauge transformations) must be
performed.

The global aspects of (5.10) are even more interesting when the coupling of the formulations to
background fields is considered.  In the $d=10$ membrane action obtained by dimensional reduction
of the $d=11$ membrane theory the local 2-form $B$ of the NS - NS sector, couples to the current
$\epsilon^{ijk} \partial_k \varphi$.  The coupling is a topological one. Assuming we are in the
euclidean worldvolume formulation of the theory, the coupling admits sources $B$ which are locally
2-forms but globally associated to nontrivial higher order $U(1)$ bundles.  The reformulation of
the action in terms of 1-forms $L_{1i}$ and constraints (5.3) and (5.4) follows as in
(5.7)-(5.10), by changing in (5.10) $f_{i}(A)$ by \begin{equation} \hat{f}_i(A)\equiv
\epsilon_{imn}(F^{mn}(A)+B^{mn}) \end{equation} where only the bosonic sector is considered. There
is an interesting change in procedure, however, arising from the nontrivial transitions of $B$.
The result is that $F^{mn}(A)$ must have also nontrivial transitions that compensate the ones of
$B$. We have in the intersection of two opens $U^\prime \cap U \neq \phi$ where a nontrivial
transition takes place \begin{eqnarray} B^{\prime} & = & B+d\eta \nonumber \\ F^{\prime} & = &
F-d\eta \end{eqnarray} which imply \begin{equation} A^\prime  = A-\eta. \end{equation}

This new transition for the connection 1-form A arises naturally in the topological field actions
introduced in {\cite{re:Martin}} to describe a gauge principle from which the Witten-Donaldson and
Seiberg-Witten invariants may be obtained as correlation functions of the corresponding BRST
invariant effective action. The most appropriate theory, however, where the nontrivial p-form
connections are expected to have  relevant non perturbative effects is the $d=11$ 5-brane. It has
been conjectured {\cite{re:Town}} that the $d=11$ 5-brane action is given by \begin{equation} S=
-\frac{1}{2}\int_{X} d^{6}\xi \sqrt{-\gamma}[\gamma^{ij}\partial_{i}x^{M}
\partial_{j}x^{N}\eta_{MN}+\frac{1}{2}\gamma^{il}\gamma^{jm}\gamma^{kn}F_{ijk}F_{lmn}-4]
\end{equation} where $F=dA$ is the self dual 3-form field strength of a local 2-form potential
$A$. We are just in the case (3.22), (3.23) discussed in section 3. There is a very rich
geometrical structure associated to this action with non perturbative effects related to the non
trivial higher order line bundles. The $d=11$ 5-brane has been also interpreted {\cite{re:Town}}
as a Dirichlet-brane of an open supermembrane, with boundary in the 5-brane worldvolume described
by a new six-dimensional superstring theory previously conjectured by \cite{re:Witten3}.  We
expect that these intrinsic non-perturbative effects should be realized naturally over non-trivial
higher order bundles . \setcounter{equation}{0} \section{Conclusions}

We found a new geometrical structure - higher order U(1) bundles- allowing a global extension of
duality transformations in quantum field theory. The intrinsic geometrical object living in these
higher order U(1) bundles are local p-forms with non trivial transitions which, in particular, are
the natural antisymmetric fields defining the D-brane actions, giving a complete topological
interpretation to their quantized charges.

The approach incorporates to the duality scheme a global constraint containing the relevant
physical parameters as coupling constants associated to the interaction of the p-forms to the
underlying p-branes,or the radius of compactification of the superstring or supermembrane. This
dependence becomes relevant in proving quantum equivalence between dual string
and membrane theories. In section five we presented an improvement ,including global aspects,of the equivalence
between the covariant $d=11$ supermembrane action with one coordinate compactified on $S^1$ and
the fully $d=10$ Lorentz covariant worldvolume action for the $d=10$ IIA Dirichlet supermembrane.

The nature of these p-form fields with non trivial transitions as well as their extension to non
abelian structure groups, will be analysed in forthcoming articles.

{\it Acknowledgements}
We are grateful to E. Planchart and L. Recht for very helpful suggestions and
discussions.

%\end{document}


\vskip 1.cm



\begin{thebibliography}{99}


\bibitem{re:Witten} N. Seiberg and E. Witten, Nucl. Phys. {\bf B426} (1994) 19,{\bf B431} (1994)
484. \bibitem{re:Ferrara} S. Ferrara, J. Scherk and B. Zumino, Nucl. Phys. {\bf B121} (1977) 393;
E. Cremmer, S. Ferrara and J. Scherk, Phys. Lett. {\bf B74} (1978) 61; A. Ceresole, R. D'Auria and
S. Ferrara, Phys. Lett.{\bf B339} (1994) 71; A. Sen, Int.J.Mod.Phys {\bf A9}(1994) 3707;
J.H.Schwarz and A. Sen, Nucl. Phys. {\bf B411} (1994) 35; I. Martin and A. Restuccia, Phys. Lett.
{\bf B323} (1994) 311; M.J.Duff and J.X.Lu,  Nucl. Phys. {\bf B426} (1994) 301; J.H. Schwarz,
Lett. Math. Phys. {\bf 34} (1995) 309; M. Duff, Nucl. Phys. {\bf B442} (1995) 47; C. Hull and P.
Townsend,  Nucl. Phys. {\bf B438} (1995) 109; E. Witten, Nucl. Phys. {\bf B443} (1995) 85; A.
Ceresole, R. D'Auria, S. Ferrara and A. van Proeyen,  Nucl. Phys. {\bf B444} (1995) 92; E. Witten,
hep-th /9507121. S. Kachru, A. Klemm, W. Lerche, P. Mayr and C. Vafa,   Nucl. Phys. {\bf B459}
(1996) 537 ; J. Stephany, hep-th/9605074. 
\bibitem{re:Steenrod} S. Eilenberg and N. Steenrod,
Foundations of Algebraic Topology, Princeton, New Jersey, Princeton University Press (1964).
\bibitem{re:Mario} F.Cachazo, M. Caicedo and A. Restuccia, preprint USB-Dec-1996.
\bibitem{re:weil} B. Kostant, Lectures Notes in Mathematics, Berlin, Springer (1970) p.133.
\bibitem{re:Witten2} E. Witten, hep-th/9505186; E. Verlinde, Nucl.Phys. {\bf
B455} (1995) 211 ; Y. Lozano, Phys. Lett. {\bf B364} (1995) 19; J.L.F. Barb\'
on, Nucl. Phys. {\bf B452} (1995) 313; A. Kehagias, hep-th/9508159. 
\bibitem{re:Town} P.K. Townsend, hep-th/9512062
\bibitem{re: x6} R. G. Leigh, Mod. Phys. Lett.{\bf A4} (1989) 2767. \bibitem{re: x7} C.
Schmidhuber, hep-th/9601003. \bibitem{re: Howe} P. Howe and R.W. Tucker, J. Phys.{\bf A10} (1977)
155; J. Math. Phys. {\bf 19} (1978) 981. \bibitem{re:Martin} R. Gianvittorio, I. Martin and A.
Restuccia, Class. Q. Gravity, {\bf 13} (1996) 2887. \bibitem{re:Witten3} E. Witten, hep-th/9507012




\end{thebibliography}
\end{document}